\documentclass[lettersize,journal]{IEEEtran}

\usepackage{fontawesome5}

\usepackage[colorlinks,urlcolor=blue,linkcolor=blue,citecolor=blue]{hyperref}
\usepackage{amsmath,amsfonts}
\usepackage{algorithmic}
\usepackage{algorithm}
\usepackage{array}
\usepackage[caption=false,font=normalsize,labelfont=sf,textfont=sf]{subfig}
\usepackage{textcomp}
\usepackage{stfloats}
\usepackage{url}
\usepackage{verbatim}
\usepackage{graphicx}
\usepackage{cite}

\usepackage{float}
\usepackage[table,xcdraw]{xcolor}
\usepackage{tcolorbox}
\usepackage{framed}
\usepackage{mdframed}
\usepackage{pgfplots}
\usepackage{xcolor}
\usepackage[utf8]{inputenc}
\pgfplotsset{compat=1.18} 
\definecolor{applegreen}{rgb}{0.55, 0.71, 0.0}
\definecolor{orange-red}{rgb}{1.0, 0.27, 0.0}
\definecolor{darkscarlet}{rgb}{0.34, 0.01, 0.1}
\usepackage{caption}
\usepackage{subcaption}
\usepackage{tikz}
\usepackage{url}

\newcommand{\llama}{\textsc{Llama-3.2-3B}}
\newcommand{\phimodel}{\textsc{Phi-4-14B}}

\newcommand{\ominihigh}{\textsc{o3-mini-high}}
\newcommand{\gemmaThree}{\textsc{Gemma-3-12B}}

\begin{document}

\title{Code Generation with Small Language Models: A Codeforces-Based Study}

\author{
    \IEEEauthorblockN{
        Débora Souza\IEEEauthorrefmark{3}, Rohit Gheyi\IEEEauthorrefmark{3}, Lucas Albuquerque\IEEEauthorrefmark{3}, Gustavo Soares\IEEEauthorrefmark{4}, Márcio Ribeiro\IEEEauthorrefmark{2}\\
    }
    \IEEEauthorblockA{
        \IEEEauthorrefmark{3}Federal University of Campina Grande, Brazil\\ \{deborasouza,lucas.albuquerque\}@copin.ufcg.edu.br and rohit@dsc.ufcg.edu.br\\
    }
    \IEEEauthorblockA{
        \IEEEauthorrefmark{4}Microsoft, USA\\gustavo.soares@microsoft.com\\
    }    
    \IEEEauthorblockA{
        \IEEEauthorrefmark{2}Federal University of Alagoas, Brazil\\ marcio@ic.ufal.br\\
    }
}

\markboth{Journal of \LaTeX\ Class Files,~Vol.~14, No.~8, August~2021}%
{Shell \MakeLowercase{\textit{et al.}}: A Sample Article Using IEEEtran.cls for IEEE Journals}

\maketitle

\begin{abstract}
Large Language Models (LLMs) demonstrate capabilities in code generation, potentially boosting developer productivity. However, their adoption remains limited by high computational costs, among other factors. Small Language Models (SLMs) present a lightweight alternative.
While LLMs have been evaluated on competitive programming tasks, prior work often emphasizes metrics like Elo or pass rates, neglecting failure analysis. The potential of SLMs in this space remains underexplored.
In this study, we benchmark three open SLMs—\llama{}, \gemmaThree{}, and \phimodel{}—across 280 Codeforces problems spanning Elo ratings from 800 to 2100 and covering 36 distinct topics. All models were tasked with generating Python solutions. 
\phimodel{} achieved the best SLM performance with a pass@3 of 63.6\%, nearing \ominihigh{} (86.8\%). Combining Python and C++ outputs increased \phimodel{}'s pass@6 to 73.6\%. A qualitative analysis revealed some failures stemmed from minor implementation issues rather than reasoning flaws.
\end{abstract}


\section{Introduction}

Recent advances in code generation area have progressed from early approaches like natural language-based synthesis~\cite{Ling2016} and grammar-based models~\cite{yin2017syntacticneuralmodelgeneralpurpose} to large pre-trained transformers such as CodeBERT~\cite{feng2020codebert} and Codex~\cite{chen2021evaluating}, which have significantly advanced the state of the art.
Despite these successes, benchmarks like HumanEval~\cite{chen2021evaluating} primarily target short, simple problems~\cite{10.1145/3597503.3639219, jiang2024surveylargelanguagemodels}, limiting their applicability to real-world scenarios. 

More recent efforts have adopted competitive programming platforms such as Codeforces~\cite{openai2025competitiveprogramminglargereasoning, li2022competition}, but often emphasize high-level metrics (e.g., Elo score, pass rate) without deeper analysis of problem diversity, model behavior, or failure modes. Additionally, evaluations based on offline datasets—such as those used by AlphaCode~\cite{li2022competition}—may overlook key constraints like runtime and memory usage.

While Large Language Models (LLMs) achieve strong
performance ~\cite{jiang2024surveylargelanguagemodels,zhao2024surveylargelanguagemodels,huynh2025largelanguagemodelscode}, their high computational demands,
energy usage, and security risks—such as data leakage and
intellectual property exposure~\cite{10.1145/3712001}—can hinder adoption.
In contrast, Small Language Models (SLMs) present a promising alternative: with fewer parameters, they offer lower latency,
easier deployment, and greater adaptability for fine-tuning in
constrained environments~\cite{wang2024comprehensivesurveysmalllanguage}. However, their capabilities in
domains like competitive programming remain underexplored.

In this study, we evaluate the performance of three small models—\llama{}, \gemmaThree{}, and \phimodel{}—on a code generation task over 280 Codeforces problems ranging from Elo 800 to 2100 and covering 36 topics. All models were tasked with generating Python code, with three submissions per problem, evaluated using the Codeforces judge for correctness, runtime, and memory usage. \phimodel{} achieved the highest performance, with a pass@3~\cite{chen2021evaluating} of 63.6\%. It was particularly effective on problems rated 800–1500 and in topics such as Strings (75.9\%), Sorting (76.3\%), and Mathematics (68.6\%). It also demonstrated high semantic consistency~\cite{ouyang2025empirical}, solving 77.5\% of problems correctly within three attempts. In contrast, other SLMs, \llama{}, \gemmaThree{}—performed significantly lower, each scoring below 19.6\%.

We also evaluated \phimodel{} on C++ code generation and found that combining outputs from both Python and C++ increased its aggregated pass@6 to 73.6\%. For comparison, \ominihigh{} achieved a pass@3 of 86.8\%. 
A manual analysis of \phimodel{}’s incorrect outputs, supported by an expert in competitive programming, revealed that some failures were caused by minor implementation issues, or language-specific limitations, rather than fundamental reasoning flaws. These results reinforce the practical potential of small, open models as efficient and reliable alternatives for code generation, approaching the performance of state-of-the-art proprietary solutions. All artifacts are publicly available online.\footnote{\url{https://zenodo.org/records/17159912}}

\section{Methodology}

The goal of this study is to evaluate the effectiveness of three language models—\llama{}, \gemmaThree{}, and \phimodel{}—in solving programming problems. This evaluation is conducted from the perspective of researchers, specifically within the domain of competitive programming. To achieve this objective, we address the following research questions:

\begin{enumerate}
    \item[\textbf{RQ$_{1}$}] \textbf{To what extent open models as \llama{}, \gemmaThree{}, and \phimodel{} can answer programming assignments?} To answer this question, the correct and incorrect responses provided by the Codeforces platform will be counted.

    \item[\textbf{RQ$_{2}$}] \textbf{What types of errors are most common in the responses generated by the SLMs?} Investigate the most frequent failures, such as syntactic errors, incorrect logic, or failure to meet problem constraints.

    \item[\textbf{RQ$_{3}$}] \textbf{How does the performance of SLMs vary across different programming topics?} Examine the SLMs' performance on problems from various topics, such as graph algorithms, dynamic programming, sorting, or string manipulation.
\end{enumerate}

\textbf{Models Selection.} The models were selected based on their potential in code-centric tasks~\cite{abdin2024phi4technicalreport}. Additionally, they represent a diverse set of architectural designs and development teams. In this work, we included models explicitly labeled as ``small'' by their developers, with fewer than 14 billion parameters—significantly smaller than proprietary state-of-the-art models from the GPT and Gemini families.

\textbf{Models Setup.}
We used default parameters from the LangChain Ollama API (temperature 0.8 unless specified). The evaluated models were \phimodel{} (14.7B parameters), \llama{} (3.2B), and \gemmaThree{} (12.2B, temperature 1), all with Q4\_K\_M quantization. An exploratory test with \phimodel{} showed that temperature 0.4 performed better, suggesting that default settings may not be optimal even for high-performing models.

\textbf{Codeforces Platform Selection.} We selected problems from the Codeforces, a widely used platform in education, technical interviews, and algorithmic competitions, known for its diverse and challenging problem sets. Its broad range of topics and difficulty levels makes it a strong benchmark for assessing the reasoning and code-generation capabilities of language models~\cite{hendrycks2021measuringcodingchallengecompetence, li2022competition, openai2025competitiveprogramminglargereasoning}.
Solving Codeforces problems requires models to comprehend complex natural language descriptions, apply suitable algorithms and data structures, and produce syntactically and semantically correct implementations. Submissions are rigorously evaluated using hidden test suites that assess correctness, edge case handling, execution time, and memory usage~\cite{li2022competition}.
Although the focus is on Codeforces programming problems, many of these tasks reflect real-world scenarios encountered by software engineers, involving the interpretation of requirement documents, the use of appropriate data structures, and the application of algorithms commonly used in day-to-day practice.

\textbf{Dataset.} To assess the performance of the models, we curated a dataset with 280 programming problems from the Codeforces API. Codeforces ranks problems using an ELO-based system, with ratings ranging from 800 to over 3000. We focused on problems rated between 800 and 2100, as fewer than 2\% of active Codeforces users had ratings above 2100 at the time of writing. For each rating level within this range, we selected 20 problems based on submission popularity as of December 15, 2024. Problem statements ranged from 111 to 1,117 tokens, with an average length of 432. 

The dataset spans 36 topics. The most frequent tags are Implementation (91 problems), Math (86), Dynamic Programming (81), Greedy (75), and Brute Force (50). Implementation problems are typically conceptually simple but require intricate coding. 
As an example, next we show part of the ``Watermelon'' problem\footnote{\url{https://codeforces.com/problemset/problem/4/A}} illustrating the typical structure of a Codeforces task, which includes a statement, input/output format, examples, and an optional note.

\begin{tcolorbox}[colframe=black!70, colback=gray!10, arc=4mm, title=Problem Watermelon]
\footnotesize
\textbf{Description} \\
    One hot summer day Pete and his friend Billy decided to buy a watermelon. [...] Pete and Billy are great fans of even numbers, that's why they want to divide the watermelon in such a way that each of the two parts weighs even number of kilos, at the same time it is not obligatory that the parts are equal. [...] For sure, each of them should get a part of positive weight.

\textbf{Input} \\
The first (and the only) input line contains integer number \( w \) \( (1 \leq w \leq 100) \) — the weight of the watermelon bought by the boys.

\textbf{Output} \\
Print YES, if the boys can divide the watermelon into two parts, each of them weighing even number of kilos; and NO in the opposite case.

\textbf{Examples} \\
Input 8
Output YES

\textbf{Note} \\
For example, the boys can divide the watermelon into two parts of 2 and 6 kilos respectively (another variant — two parts of 4 and 4 kilos).
\end{tcolorbox}

\textbf{Prompt Design.} To improve solution accuracy, we designed prompts with essential contextual information, including a persona definition, language-specific instructions (Python), and clear guidelines on input/output handling. Additionally, the prompt describes the structure of the Codeforces problem statement, ensuring the model interprets it correctly. Together with the actual problem, these components provide a well-rounded foundation to guide the code generation process.

\begin{mdframed}[
    backgroundcolor=cyan!10, 
    linewidth=1pt,           
    linecolor=black,          
    innerleftmargin=5pt,     
    innerrightmargin=5pt,
]
\footnotesize
    You are a highly skilled competitive programmer with 15 years of experience in the field. \\
    Your objective is to analyze the following problem statement and to produce a Python code solution that adheres to the requirements.\\
    Guidelines for the solution: deliver only the Python code; Ensure the solution reads input via standard input and produces outputs results via standard output; If the solution requires defining a function, ensure it is executed within the code; Avoid adding explanations, comments, or unnecessary text. \\
    The Problem Statement includes a detailed description, input and output format, and examples to clarify requirements.\\
    \{\textit{Problem Statement}\}
\end{mdframed}

\textbf{Model Output.} The prompt above explicitly instructs the model to respond with Python code only, avoiding explanations, comments, or unnecessary text. However, it is well known that language models may still include additional text around the code despite such instructions~\cite{macedo2024exploring}. Therefore, post-processing was applied using the following regular expression to extract only the generated code segment:  \texttt{```python\textbackslash n?(.*?)(?:\textbackslash n?{```}|\$)}.

\textbf{Procedure.} We developed an automated pipeline to evaluate all the models.  
Each model submitted three solutions per problem.
All executions were performed locally in March 2025 using the Ollama framework on a NVIDIA RTX 3060 GPU (12GB VRAM), with default parameters from the LangChain Ollama API to ensure consistency across models.
The process was fully automated: problem statements were extracted, prompts generated, code produced by the models, and solutions submitted to Codeforces.

\section{Evaluation}
\label{sec:eval}

This section presents the results for each research question and provides a discussion of the findings.

\subsection{RQ$_1$. Models Performance}

We evaluate models using the \textit{pass@k} metric~\cite{chen2021evaluating}, which estimates the probability that at least one of the top-\textit{k} generated solutions is correct. As shown in Table~\ref{tab:unified_ieee}, \phimodel{} (Python) outperforms all other models, achieving 48.3\% at pass@1 and 63.6\% at pass@3. \gemmaThree{} follows with 16.9\% and 19.6\%, respectively, outperforming \llama{}, which stay below 11\% at pass@3, suggesting that smaller models struggle with more complex programming tasks.
We also analyzed the length of generated outputs. As shown in Table~\ref{tab:unified_ieee}, \phimodel{} and \gemmaThree{} produce longer solutions, averaging 24.3 and 27.7 lines of code, with over 200 tokens on average. Notably, \phimodel{} also produces the longest individual solution (1,163 tokens).

\begin{table}[htbp]
\caption{Evaluation of \llama{} (Lla), \gemmaThree{} (Gem3), and \phimodel{} (Phi) across pass@k, Codeforces topics, and semantic consistency.}
\label{tab:unified_ieee}
\centering
\begin{tabular}{|l|c|c|c|}
\hline
\rowcolor[HTML]{000000}
\textbf{\color[HTML]{FFFFFF}} & \textbf{\color[HTML]{FFFFFF}Lla} & \textbf{\color[HTML]{FFFFFF}Gem3} & \textbf{\color[HTML]{FFFFFF}Phi} \\ \hline
\multicolumn{4}{|l|}{\textbf{Accuracy (pass@k)}} \\ \hline
pass@1 & 7.0\% & 16.9\% & 48.3\% \\ \hline
pass@2 & 9.6\% & 18.6\% & 58.8\% \\ \hline
pass@3 & 11.1\% & 19.6\% & 63.6\% \\ \hline
\multicolumn{4}{|l|}{\textbf{Topical Accuracy (pass@3)}} \\ \hline
Implementation & 17.6\% & 34.1\% & 83.5\% \\ \hline
Sortings & 7.9\% & 10.5\% & 76.3\% \\ \hline
Strings & 34.5\% & 51.7\% & 75.9\% \\ \hline
Two Pointers & 0.0\% & 4.3\% & 52.2\% \\ \hline
Math & 8.1\% & 19.8\% & 68.6\% \\ \hline
Brute Force & 14.0\% & 32.0\% & 68.0\% \\ \hline
Combinatorics & 0.0\% & 11.1\% & 66.7\% \\ \hline
Greedy & 10.7\% & 12.0\% & 65.3\% \\ \hline
Binary Search & 0.0\% & 5.1\% & 61.5\% \\ \hline
Constructive Alg. & 3.0\% & 3.0\% & 57.6\% \\ \hline
\multicolumn{4}{|l|}{\textbf{Semantic Consistency}} \\ \hline
Score & 61.3\% & 85.4\% & 77.5\% \\ \hline
\multicolumn{4}{|l|}{\textbf{Generated Code Length}} \\ \hline
LOC\_MEAN & 19.5 & 27.7 & 24.3 \\ \hline
LOC\_MAX & 81 & 99 & 106 \\ \hline
MEAN\_TOKEN & 164.8 & 210.2 & 204.7 \\ \hline
MAX\_TOKEN & 752 & 762 & 1,163 \\ \hline
\multicolumn{4}{|l|}{\textbf{Phi-4 in C++ (pass@1 / 2 / 3)}} \\ \hline
\phimodel{} (C++)     & \multicolumn{3}{c|}{47.7\% / 57.3\% / 61.1\%} \\ \hline
\multicolumn{4}{|l|}{\textbf{Accuracy of Additional Models (pass@1 / 2 / 3)}} \\ \hline
\ominihigh{}        & \multicolumn{3}{c|}{83.3\% / 85.7\% / 86.8\%} \\ \hline
\end{tabular}
\end{table}

\subsection{RQ$_2$. Types of Errors}

When submitting a solution to a Codeforces problem, verdicts other than Accepted—such as Wrong Answer, Compilation Error, Time Limit Exceeded (TLE), Memory Limit Exceeded (MLE), and Runtime Error—indicate specific types of failures. TLE and MLE indicate that the submitted code fails to meet the time and memory constraints, respectively. Figure~\ref{fig:errors_auto} presents the distribution of these errors across models. Wrong Answer is the most frequent error across all models, indicating incorrect outputs for one or more test cases. Runtime Error, often caused by invalid memory access or uninitialized variables, is the second most common failure, particularly affecting \llama{} (27\%). TLE is also notable, especially for \gemmaThree{} (39.3\%), suggesting inefficiencies in the generated algorithms.

\begin{figure}[ht]
    \centering
    \begin{tikzpicture}
    \begin{axis}[
        ybar,
        bar width=10pt,
        width=0.52\textwidth,
        height=4cm,
        ymin=0,
        ymax=70,
        ylabel={Error Percentage (\%)},
        xlabel style={
            at={(axis description cs:0.5,-0.19)},
            anchor=north
        },
        symbolic x coords={
            Llama 3.2 3B,
            Gemma 3 12B,
            Phi-4 14B,
        },
        xtick=data,
        enlarge x limits=0.23,
        legend style={
            at={(0.5,1.02)},
            anchor=south,
            legend columns=3,
            font=\footnotesize,
            draw=none,
            fill=none
        },
        nodes near coords,        
        nodes near coords align={vertical},
        grid=major,
        grid style={dashed,gray!30},
        legend image post style={draw=none},
        legend image code/.code={
            \draw[#1, draw=none] (0cm,-0.1cm) rectangle (0.3cm,0.1cm);
        },
        nodes near coords style={text=black},
    ]
        \addplot+[draw=none, fill=red!80] coordinates {
            (Llama 3.2 3B,58.5)
            (Gemma 3 12B,53.9)
            (Phi-4 14B,33.9)
        };
        \addlegendentry{Wrong Answer}

        \addplot+[draw=none, fill=orange!80] coordinates {
            (Llama 3.2 3B,27)
            (Gemma 3 12B,3.6)
            (Phi-4 14B,6.4)
        };
        \addlegendentry{Runtime Error}

        \addplot+[draw=none, fill=blue!70] coordinates {
            (Llama 3.2 3B,5.0)
            (Gemma 3 12B,39.3)
            (Phi-4 14B,9.4)
        };
        \addlegendentry{Time Limit Exceeded}

        \addplot+[draw=none, fill=pink!70] coordinates {
            (Llama 3.2 3B,2.5)
            (Gemma 3 12B,1.4)
            (Phi-4 14B,0.2)
        };
        \addlegendentry{Compilation Error}

        \addplot+[draw=none, fill=purple!70] coordinates {
            (Llama 3.2 3B,0.2)
            (Gemma 3 12B,1.9)
            (Phi-4 14B,1.7)
        };
        \addlegendentry{Memory Limit Exceeded}
    \end{axis}
    \end{tikzpicture}
    \captionsetup{format=plain}
    \caption{Error rate across 840 submissions (280 problems × 3).}
    \label{fig:errors_auto}
\end{figure}

\subsection{RQ$_3$. Model Effectiveness by Topic}

Each Codeforces problem is associated with one or more topics. Table~\ref{tab:unified_ieee} summarizes model performance across 10 of the 36 topics considered in our study, including Math and Greedy algorithms—each requiring distinct forms of algorithmic reasoning.
\phimodel{} outperforms all other models across every topic, achieving the highest pass@3 accuracy. Its performance is particularly strong in \textit{Implementation} (83.5\%), \textit{Sortings} (76.3\%), and \textit{Strings} (75.9\%), with some gaps over the second-best model. It also leads in more complex categories like \textit{Binary Search} (61.5\%) and \textit{Brute Force} (68\%), demonstrating generalization across a range of problem types. \gemmaThree{} consistently improves over \llama{}, especially in \textit{Strings} (51.7\%) and \textit{Brute Force} (32\%), suggesting that moderate increases in model size or training can yield measurable gains.
\llama{} struggle on more demanding topics. For instance, in \textit{Combinatorics}, \llama{} scores 0\%, while \phimodel{} achieves 66.7\%. A similar trend appears in \textit{Binary Search}, \textit{Constructive Algorithms}, and \textit{Two Pointers}, where only \phimodel{} surpasses 50\% accuracy.

\subsection{Performance Breakdown by Difficulty}
\label{sec:rating}

\phimodel{} (Python) consistently outperforms all open-SLMs across nearly all difficulty levels. However, all SLMs show a  performance decline as problem difficulty increases, highlighting their limitations with complex tasks and relative strength on easier ones. Each SLM exhibits different performance profiles. \llama{} performs reasonably well up to rating 900, with only sporadic success beyond that. \gemmaThree{} extends that range to around 1000, although with noticeable drops in accuracy after 1200. \phimodel{} spans the full range up to 2100, but shows a decline beyond 1500, indicating challenges with higher complexity. 

\begin{figure*}[ht]
\centering
\small
\begin{tikzpicture}
\begin{axis}[
  width=0.9\textwidth,
  height=6.1cm,
  xlabel={Codeforces Rating (K)},
  xlabel style={
    at={(axis description cs:0.5,-0.17)},
    anchor=south
  },
  ylabel style={
    at={(axis description cs:-0.07,0.5)},
    anchor=south
  },
  xmin=800, xmax=2100,
  ymin=0, ymax=1.05,
  ytick={0,0.2,0.4,0.6,0.8,1.0},
  yticklabels={0\%, 20\%, 40\%, 60\%, 80\%, 100\%},
  ylabel={Correct answers (\%)},
  xtick={800,900,1000,1100,1200,1300,1400,1500,1600,1700,1800,1900,2000,2100},
  xticklabels={0.8, 0.9, 1, 1.1, 1.2, 1.3, 1.4, 1.5, 1.6, 1.7, 1.8, 1.9, 2, 2.1},
  xticklabel style={font=\footnotesize},
  legend columns=3,
  legend style={
    font=\footnotesize,
    column sep=1ex,
    draw=none,
    at={(0.5,-0.15)},
    anchor=north
  },
  ymajorgrids=true,
  grid style=dashed
]

\addplot[color=orange, mark=square*, thick, mark size=1.5pt] coordinates {
  (800,1.00) (900,0.95) (1000,1.00) (1100,1.00)
  (1200,0.85) (1300,1.00) (1400,0.75) (1500,0.65)
  (1600,0.45) (1700,0.35) (1800,0.15) (1900,0.30)
  (2000,0.30) (2100,0.15)
};
\addlegendentry{\phimodel{} (Python)}

\addplot[color=green!60!black, mark=x, thick, mark size=1.5pt] coordinates {
  (800,1.00) (900,0.45) (1000,0.55) (1100,0.10)
  (1200,0.30) (1300,0.10) (1400,0.10) (1500,0.10)
  (1600,0.00) (1700,0.05) (1800,0.00) (1900,0.00)
  (2000,0.00) (2100,0.00)
};
\addlegendentry{\gemmaThree{}}

\addplot[color=blue, mark=*, thick, mark size=1.5pt] coordinates {
  (800,0.65) (900,0.40) (1000,0.25) (1100,0.10)
  (1200,0.00) (1300,0.05) (1400,0.00) (1500,0.05)
  (1600,0.00) (1700,0.05) (1800,0.00) (1900,0.00)
  (2000,0.00) (2100,0.00)
};
\addlegendentry{\llama{}}

\addplot[color=black, mark=triangle*, thick, mark size=1.5pt] coordinates {
  (800,1.00) (900,1.00) (1000,1.00) (1100,1.00)
  (1200,1.00) (1300,1.00) (1400,0.90) (1500,0.95)
  (1600,1.00) (1700,0.80) (1800,0.65) (1900,0.70)
  (2000,0.65) (2100,0.50)
};
\addlegendentry{\ominihigh{}}

\addplot[color=magenta, mark=o, thick, mark size=1.5pt] coordinates {
  (800,1.00) (900,0.90) (1000,0.90) (1100,0.95)
  (1200,0.60) (1300,0.80) (1400,0.60) (1500,0.70)
  (1600,0.45) (1700,0.35) (1800,0.40) (1900,0.35)
  (2000,0.35) (2100,0.20)
};
\addlegendentry{\phimodel{} (C++)}

\addplot[color=red, mark=o, thick, mark size=1.5pt] coordinates {
  (800,1.00) (900,1.00) (1000,1.00) (1100,1.00)
  (1200,0.90) (1300,1.00) (1400,0.80) (1500,0.80)
  (1600,0.55) (1700,0.60) (1800,0.45) (1900,0.60)
  (2000,0.40) (2100,0.20)
};
\addlegendentry{Phi-4 (C++ and Python)}

\end{axis}
\end{tikzpicture}

\caption{
Performance by Codeforces difficulty (800–2100). Values show correct solutions (out of 20) with pass@3, except \textit{\phimodel{} (C++/Python)} using pass@6.
}
\label{fig:rating_comparison_one_column}
\end{figure*}

\subsection{Consistency Analysis}

While \textit{pass@k} measures how often a model produces at least one correct solution, it does not capture consistency across multiple attempts. To address this, we introduce Semantic Consistency (SC), which evaluates how reliably a model reproduces correct outputs in repeated submissions~\cite{ouyang2025empirical}. This metric is particularly important for code generation, where consistent correctness matters more than isolated success. As shown in Table~\ref{tab:unified_ieee}, all models achieved SC above 60\%, with \gemmaThree{} leading despite lower overall accuracy. \phimodel{} also showed strong consistency (77.5\%), indicating that once the model solves a problem, it tends to do so repeatedly—a sign of learned, stable behavior rather than random chance.

\subsection{\phimodel{}: C++}
\label{sec:other_languages}

To investigate whether the choice of programming language influences model performance, we evaluated \phimodel{}, which has the best performance in our evaluation, using both Python and C++. C++ is widely adopted in competitive programming due to its performance efficiency, particularly in time-critical problems. In this experiment, we kept the prompting strategy, modifying only the target language. The combined results reported represent the union of problems solved by either language, simulating an ideal scenario where solutions in both languages can be leveraged to maximize coverage.

The results show that \phimodel{} achieves comparable accuracy in both languages: 63.6\% pass@3 in Python and 61.1\% in C++. However, the sets of problems solved in each language differ around 20\%. The Python version solved 35 problems not solved in C++, while the C++ version succeeded on 28 problems not handled in Python. The combined results—union of problems solved in either language—represent an upper-bound scenario where multi-language solutions maximize coverage (see Figure~\ref{fig:rating_comparison_one_column}) In this case, \phimodel{} solves 206 out of 280 problems—achieving an aggregated pass@2 of 63.6\%, pass@4 of 70.4\%, and pass@6 of 73.6\%. 

Of the 28 problems uniquely solved in C++, 25 are rated above 1500 and primarily fall under categories such as Dynamic Programming, Math, Greedy, Brute Force, and Graphs. Python failures on these problems were often due to Time Limit Exceeded (12), Wrong Answers (18), or Runtime Errors (9), suggesting that differences are driven by computational constraints and problem complexity rather than the language itself.

Although developers may not always have the freedom to choose the programming language for a given task—due to platform restrictions, legacy systems, or deployment environments—multi-language generation remains a valuable strategy in several contexts. For instance, in latency-critical scenarios such as AWS Lambda, developers may prefer deploying solutions in more performant languages like C++ to reduce execution time. Recent work such as InterTrans~\cite{macedo2024intertrans} also highlights how leveraging multiple target languages can improve coverage and effectiveness in code generation pipelines, reinforcing the practical relevance of this approach.

\subsection{Larger Models}
\label{sec:performance-comparison}

To understand how far SLMs are from the best-performing models, we analyzed \ominihigh{}, which, as shown in Table~\ref{tab:unified_ieee}, achieved the highest overall accuracy with a pass@3 of 86.8\% on the same dataset. \phimodel{} successfully solved five problems that were not answered by \ominihigh{}. These problems had Elo ratings of 1400, 1500, 1900 (2x) and 2100. These findings confirm that even state-of-the-art proprietary models struggle with high-difficulty competitive programming tasks.

However, when aggregating \phimodel{}'s results across Python and C++ (Section~\ref{sec:other_languages}), its pass@3 increases to 73.6\%, narrowing the gap to \ominihigh{}. This highlights the potential of open models to achieve competitive performance through multi-language strategies. Both models follow a similar trend: strong accuracy on easier problems with a gradual decline as difficulty increases (Figure~\ref{fig:rating_comparison_one_column}).
Moreover, \phimodel{} is open, cost-effective, and runs on consumer-grade hardware, making it an attractive alternative for a wide range of applications. 

\subsection{Metamorphic Testing}

To assess data contamination~\cite{carlini2023extracting} in our benchmark, we conducted metamorphic testing on 14 problems with at least one submission accepted—one per difficulty level—using \phimodel{} and Python language, the best-performing model. This process involved altering problem statements through minor edits such as renaming variables, replacing verbs, and using synonyms, without changing their core semantics. Before transformation, \phimodel{} achieved pass@1/2/3 scores of
78.57\%, 92.86\%, and 100\%. After transformation, the scores were 69.05\%, 80.95\%, and 85.71\%, respectively. Besides the small drop in performance, these values still suggest no evidence of data contamination.

\subsection{Qualitative Analysis}
\label{o3-mini}

We conducted a qualitative analysis of the errors made by \phimodel{}. One of the authors, a competitive programmer with over seven years of experience, has reached a peak ELO rating of 2245 on Codeforces. He conducted a manual review of a sample of 21 ($\texttildelow20\%$) problems that \phimodel{} failed to solve. Based on this analysis, we identified four primary causes of failure: near-correct solutions with minor mistakes, algorithmically correct solutions that exceeded time limits, incorrect solutions with relevant direction, and fundamentally flawed implementations.

The first category of failures involved near-correct solutions that required only minor adjustments to succeed. In five of the reviewed problems, \phimodel{} produced almost correct code, needing simple fixes such as correcting variable initialization (Problem 1343-C), handling edge cases (Problem 1399-C), or adjusting state transitions in dynamic programming (Problem 698-A).
Three additional problems required more modifications. In Problem 489-C, multiple assignment errors and output formatting issues had to be corrected. Problem 1-B involved a minor logical issue, which required adding a loop to properly parse the input. In Problem 161-D, the Python solution was logically sound but inefficient; performance improvements were necessary, including replacing dictionaries with lists and rewriting recursive functions as iterative ones to meet time constraints.

In the second category, three problems featured conceptually correct algorithms that failed due to time constraints. When we prompted \phimodel{} to rewrite these solutions in C++—preserving the logic but aiming for improved efficiency—two were successfully solved. However, Problem 1541-B (Elo 1200) remained unsolved, as it required a reduction in time complexity beyond what the model could infer.

Many of \phimodel{}'s incorrect answers were due to minor implementation issues or edge-case handling, rather than fundamental misunderstandings. These errors were often correctable with small edits. Addressing such cases systematically could bring the model's performance even closer to that of proprietary models like \ominihigh{}, reinforcing the value of open solutions in competitive programming tasks.

In 4 of the reviewed problems, \phimodel{} produced outputs that were in the right direction but still far from being correct. These solutions often included algorithmic elements aligned with the correct approach—such as identifying the right data structure or outlining the high-level logic—but lacked crucial implementation details or contained major logical errors. While the model appeared to grasp the core idea behind the problems, its execution was flawed and incomplete. As a result, these attempts were not easily repairable and required substantial rewriting to become viable.
Still, one noteworthy aspect is that this type of partially correct attempt can offer valuable insights to the user. In problems where it is particularly hard to identify a viable direction, having the model suggest an initial high-level approach—even if incomplete—can be quite helpful as a starting point for further reasoning.
In Problem 339-D (Elo 1700), the failure was attributed to incomplete problem description caused by formatting issues during automatic extraction. For example, numbers with exponents were incorrectly ignored, altering the meaning of the problem statement. As future work, we plan to enhance the text extraction process to avoid such issues.

Finally, in 5 of the 21 cases, the model produced fundamentally incorrect code. These outputs were either logically incoherent or based on an incorrect understanding of the problem statement. In some cases, the specialist reviewer noted that ``\textit{the model seems not to have understood the problem at all.}'' In all 3 attempts, the model made incorrect assumptions and attempted to solve a completely different task. To reduce costs, prioritize open models—\phimodel{} in Python, then C++—and check outputs for minor implementation errors before using expensive proprietary options like \ominihigh{}.

\subsection{Threats to Validity} 

This study is limited by the use of fixed prompts, quantization settings, models, and a dataset of 280 popular Codeforces problems, which may introduce topic and difficulty biases. Results may vary with alternative configurations or larger benchmarks. Manual error analysis by a single specialist may introduce bias and pass@k only captures success rates, not code quality aspects.

\section{Related Work}

Hou and Ji~\cite{Hou_2024} evaluated LLMs, including GPT-4, Claude 2, LLaMA 2, GPT-3.5, Gemini Ultra, Gemini Pro, and Code LLaMA, on LeetCode and GeeksforGeeks challenges across varying difficulty levels, allowing up to five attempts per problem. Sakib et al.~\cite{Sakib_2024} focused on GPT-3.5 with stricter constraints (two attempts) across three difficulty levels and diverse algorithmic topics. Their evaluation covered three difficulty levels and a diverse range of algorithmic topics, offering insights into how well LLMs handle different problem-solving paradigms.

Jain et al.\cite{jain2024livecodebenchholisticcontaminationfree} proposed LiveCodeBench, a holistic, contamination-free evaluation framework for LLMs using Codeforces problems. Continuously updated with new problems from Codeforces and other platforms, it assesses a wide range of code-related capabilities, including code generation, self-repair, execution, and test case prediction. The study evaluated 18 base and 34 instruction-tuned LLMs, uncovering insights on dataset contamination, performance comparisons, and overfitting. Complementing benchmark-based evaluations, recent agentic-driven approaches such as self-collaboration\cite{dong2024selfcollab} and self-planning~\cite{jiang2024selfplanning} improve problem-solving by structuring tasks into subtasks or plans, achieving up to 47.1\% gains over single-agent baselines. Large-scale analyses on platforms like Codeforces~\cite{hendrycks2021measuringcodingchallengecompetence} further reveal model capabilities across topics and difficulties, e.g., GPT-Neo solving around 20\% of introductory problems, highlighting incremental advances in code correctness.

While previous works have analyzed the performance of models like AlphaCode~\cite{li2022competition} and \ominihigh{}\cite{openai2025competitiveprogramminglargereasoning} on Codeforces, focusing mainly on correctness, our study instead highlights the behavior of smaller models, covering multiple topics and Elo ratings and adding a qualitative error analysis. Recent benchmarks such as LiveCodeBench Pro~\cite{livecodebenchpro2025} and Wei et al.~\cite{wei2025evaluatingimprovinglargelanguage} provide valuable insights into the limitations of large models across problem types and reasoning categories, and our work complements them by shifting the focus to SLMs with a more fine-grained evaluation.

\section{Conclusion}

This study evaluated the performance of three open SLMs—\llama{}, \gemmaThree{}, and \phimodel{}—on 280 competitive programming problems from Codeforces. \phimodel{} stood out with a pass@3 of 63.6\% and semantic consistency of 77.5\%, significantly outperforming other SLMs. In contrast, smaller models like \llama{} and \gemmaThree{} achieved pass@3 rates below 20\%. We also evaluated \phimodel{} on C++ code generation and found that combining outputs from both Python and C++ increased its aggregated pass@6 to 73.6\%. For comparison, the proprietary \ominihigh{} model achieved a pass@3 of 86.8\%.

A manual review of incorrect \phimodel{} submissions revealed that most solutions were close to correct, with errors arising from minor issues—such as unhandled edge cases or suboptimal implementations—rather than fundamental reasoning flaws. Notably, 26 problems that failed in Python were successfully solved in C++, suggesting that leveraging multiple programming languages can enhance model performance.
These results show that SLMs—particularly \phimodel{}—strike a compelling balance between performance, efficiency, and accessibility. They represent a practical alternative to large proprietary models, especially in settings with limited resources. In environments with tighter computational constraints, smaller models such as \llama{}, or \gemmaThree{} can still perform effectively on lower-difficulty problems (e.g., Codeforces ratings between 800 and 1000). Future work may explore higher-ELO problems, refined prompting strategies, alternative configurations, language-specific generation, and lightweight tuning techniques to further assess and improve model effectiveness.

As future work, we intend to include efficiency-oriented metrics such as inference time, latency, energy consumption, and cost-per-accepted solutions, which are critical for practical deployment but not assessed here. Contamination was only checked through metamorphic testing, and more rigorous approaches like time-split controls are needed. Broader baselines and ablation studies (e.g., temperature, seeds, quantization, self-repair) could also clarify configuration sensitivity, while automated pipelines for bug categorization would refine error analysis beyond the manual audit conducted in this work. 

\section*{Acknowledgments}
We thank the anonymous reviewers for their valuable feedback. This work was partially supported by CNPq (403719/2024-0, 310313/2022-8, 404825/2023-0, 443393/2023-0, 312195/2021-4), FAPESQ-PB (268/2025).

\bibliographystyle{IEEEtran}

\begin{thebibliography}{10}
\providecommand{\url}[1]{#1}
\csname url@samestyle\endcsname
\providecommand{\newblock}{\relax}
\providecommand{\bibinfo}[2]{#2}
\providecommand{\BIBentrySTDinterwordspacing}{\spaceskip=0pt\relax}
\providecommand{\BIBentryALTinterwordstretchfactor}{4}
\providecommand{\BIBentryALTinterwordspacing}{\spaceskip=\fontdimen2\font plus
\BIBentryALTinterwordstretchfactor\fontdimen3\font minus \fontdimen4\font\relax}
\providecommand{\BIBforeignlanguage}[2]{{%
\expandafter\ifx\csname l@#1\endcsname\relax
\typeout{** WARNING: IEEEtran.bst: No hyphenation pattern has been}%
\typeout{** loaded for the language `#1'. Using the pattern for}%
\typeout{** the default language instead.}%
\else
\language=\csname l@#1\endcsname
\fi
#2}}
\providecommand{\BIBdecl}{\relax}
\BIBdecl

\bibitem{macedo2024exploring}
M.~Macedo, Y.~Tian, F.~R.~Cogo, and B.~Adams, ``{Exploring the Impact of the Output Format on the Evaluation of Large Language Models for Code Translation},'' arXiv:2403.17214, 2024.


\bibitem{jiang2024selfplanning}
X.~Jiang et al., ``{Self-Planning Code Generation with Large Language Models},'' \textit{ACM Transactions on Software Engineering and Methodology}, vol.~33, no.~7, pp.~1--30, Sep. 2024.


\bibitem{Hou_2024}
W.~Hou and Z.~Ji, ``{Comparing Large Language Models and Human Programmers for Generating Programming Code},'' \textit{Advanced Science}, Wiley, Dec. 2024.

\bibitem{carlini2023extracting}
N.~Carliniet al., ``{Extracting training data from diffusion models},'' in \emph{Proc. USENIX Security Symposium}, Anaheim, CA, USA, USENIX Association, 2023, Art. no. 294, pp. 1--18.

\bibitem{jain2024livecodebenchholisticcontaminationfree}
N.~Jain et al., ``{LiveCodeBench: Holistic and Contamination Free Evaluation of Large Language Models for Code},'' arXiv:2403.07974, 2024.


\bibitem{Sakib_2024}
F.~A.~Sakib, S.~H.~Khan, and A.~H.~M.~R.~Karim, ``{Extending the Frontier of ChatGPT: Code Generation and Debugging},'' in \textit{Proc. Int. Conf. on Electrical, Computer and Energy Technologies}, pp. 1–6, 2024.

\bibitem{livecodebenchpro2025}
Z.~Zheng et al., ``{LiveCodeBench Pro: How Do Olympiad Medalists Judge LLMs in Competitive Programming?},'' arXiv:2506.11928, 2025.

\bibitem{dong2024selfcollab}
Y.~Dong, X.~Jiang, Z.~Jin, and G.~Li, ``{Self-Collaboration Code Generation via ChatGPT},'' \textit{ACM Transactions on Software Engineering and Methodology}, vol.~33, no.~7, pp.~1--38, Sep. 2024.


\bibitem{macedo2024intertrans}
M.~Macedo et al., ``{InterTrans: Leveraging Transitive Intermediate Translations to Enhance LLM‑based Code Translation},'' arXiv:2411.01063, 2024.

\bibitem{wei2025evaluatingimprovinglargelanguage}
M.~Wei et al., ``{Evaluating and Improving Large Language Models for Competitive Program Generation},'' arXiv:2506.22954, 2025.


\bibitem{Ling2016}
W.~Ling et al., ``{Latent predictor networks for code generation},'' arXiv:1603.06744, 2016.

\bibitem{yin2017syntacticneuralmodelgeneralpurpose}
P.~Yin and G.~Neubig, ``{A syntactic neural model for general-purpose code generation},'' arXiv:1704.01696, 2017.

\bibitem{feng2020codebert}
Z.~Feng et al., ``{CodeBERT}: A pre-trained model for programming and natural languages,'' arXiv:2002.08155, 2020.

\bibitem{chen2021evaluating}
M.~Chen et al., ``{Evaluating large language models trained on code},'' arXiv:2107.03374, 2021.

\bibitem{10.1145/3597503.3639219}
X.~Du et al., ``{Evaluating Large Language Models in Class-Level Code Generation},'' in \emph{International Conference on Software Engineering}, 2024, pp. 982--994.

\bibitem{jiang2024surveylargelanguagemodels}
J.~Jiang, F.~Wang, J.~Shen, S.~Kim, and S.~Kim, ``{A survey on large language models for code generation},'' arXiv:2406.00515, 2024.

\bibitem{openai2025competitiveprogramminglargereasoning}
A.~El-Kishky et al., ``{Competitive Programming with Large Reasoning Models},'' arXiv:2502.06807, 2025.

\bibitem{li2022competition}
Y.~Li et al., ``{Competition-level Code Generation with AlphaCode},'' \emph{Science}, vol. 378, no. 6624, pp. 1092--1097, 2022.

\bibitem{zhao2024surveylargelanguagemodels}
W.~X. Zhao et al., ``{A Survey of Large Language Models},'' arXiv:2303.18223, 2024.

\bibitem{huynh2025largelanguagemodelscode}
N.~Huynh and B.~Lin, ``{Large Language Models for Code Generation: A Comprehensive Survey of Challenges, Techniques, Evaluation, and Applications},'' arXiv:2503.01245, 2025.

\bibitem{10.1145/3712001}
B.~C. Das, M.~H. Amini, and Y.~Wu, ``{Security and Privacy Challenges of Large Language Models: A Survey},'' \emph{ACM Computing Surveys}, vol.~57, no.~6, pp. 1--39.

\bibitem{wang2024comprehensivesurveysmalllanguage}
F.~Wang et al., ``{A Comprehensive Survey of Small Language Models in the Era of Large Language Models: Techniques, Enhancements, Applications, Collaboration with LLMs, and Trustworthiness},'' arXiv:2411.03350, 2024.

\bibitem{abdin2024phi4technicalreport}
M.~Abdin et al., ``{Phi-4 Technical Report},'' arXiv:2412.08905, 2024.

\bibitem{hendrycks2021measuringcodingchallengecompetence}
D.~Hendrycks et al., ``{Measuring Coding Challenge Competence With APPS},'' arXiv:2105.09938, 2021.

\bibitem{ouyang2025empirical}
S.~Ouyang, J.~M. Zhang, M.~Harman, and M.~Wang, ``{An Empirical Study of the Non-Determinism of ChatGPT in Code Generation},'' \emph{ACM Transactions on Software Engineering and Methodology}, vol.~34, no.~2, pp. 1--28.

\end{thebibliography}

\vfill

\end{document}